# The User Side of AI Model Lifecycles: Evidence from the Keep4o Movement

**Yiwen Wu**
Independent Researcher
ORCID: 0009-0009-9047-8596
Email: yiwen.wu.research@proton.me

**Abstract**

AI model lifecycles are commonly understood as a series of technical and organizational processes. Yet once a model enters sustained use, subsequent changes can also affect established user practices and user value. Using the Keep4o movement around GPT-4o as a case, this study examines post-deployment AI model lifecycle issues from the user side. We collected 61,846 public original posts on X from August 2025 to March 2026 and, using a systematically developed coding framework and LLM-assisted content analysis, analyzed discussion themes, users' reasons for wanting to keep GPT-4o, and the specific claims they made.

Findings show that the Keep4o discussion extended well beyond continued access to the model itself. It covered concrete experiences of use, model behavioral characteristics and how they changed, and management issues across different stages of the model lifecycle. Reasons for keeping GPT-4o reflected interactional and relational value formed through long-term use, as well as judgments about the adequacy of replacement and the reasonableness of related decisions. The corresponding claims further reflected users' specific expectations for model lifecycle arrangements and governance. Overall, the call to "keep GPT-4o" brought together different judgments about user value and governance concerns.

These findings suggest that technical version succession does not necessarily amount to effective replacement on the user side. Post-deployment AI model lifecycle management therefore needs to consider whether established user value can be carried forward and how model changes affect actual use. This study thus provides user-side empirical evidence for AI model lifecycle management. It further shows that user experience can provide important information for identifying post-deployment impacts and should be incorporated into lifecycle evaluation and decision-making.

# 1 Introduction

As artificial intelligence (AI) becomes increasingly embedded in work and everyday life through user-facing products and services, users may develop stable expectations and practices around particular AI models. Subsequent changes to these models, including replacement or retirement, can affect established use and the conditions that support it. Post-deployment AI model lifecycle management has therefore become an important governance issue for AI services.

This study examines the controversy surrounding the lifecycle management of GPT-4o[1] and the Keep4o movement that emerged around it. GPT-4o is a native multimodal large language model (LLM) developed by OpenAI. Keep4o is a user movement that emerged as some users took to social media to express their views and organize collective action around the impacts they experienced as GPT-4o changed over time, including changes while it remained in service and its eventual retirement.

Existing research has shown that sustained human–AI interaction can generate value beyond discrete functions, that technical configurations and rules in digital services can shape users' available choices, and that the retirement of technical components can create migration and continuity challenges. Yet these issues have largely been addressed across separate bodies of literature. Systematic research remains limited on how post-deployment changes to AI models that users have used over extended periods affect their experience, and how lifecycle management should account for these impacts.

Specifically, this study addresses four research questions. First, what issues do social media expressions within the Keep4o movement primarily concern? Second, what reasons do users give for retaining GPT-4o or opposing related changes? Third, what specific claims do users make regarding changes to GPT-4o? Fourth, what responsibilities and normative issues in AI model lifecycle management are revealed by these themes, reasons, and claims?

The main contribution of this study is to examine post-deployment AI model lifecycle management from the user side. By analyzing large-scale public discourse surrounding the Keep4o movement, the study connects user value formed through sustained use to model lifecycle issues and further considers the role of user experience in lifecycle management.

# 2 Literature Review

## 2.1 Relational Value in Conversational AI Use

User value in conversational AI can develop through sustained interaction. Research on human–AI interaction (HAI) has long treated trust as an important dimension of understanding interactions between people and AI. Ueno et al. (2022), in their review of relevant models, measures, and methods, show that trust has become an important and multifaceted issue in HAI research. This line of research establishes trust as a key dimension of HAI, providing a basis for examining how users' evaluations of AI develop through interaction.

This relational value has a clear temporal dimension and develops gradually through sustained interaction. In interviews with 18 Replika users, Skjuve et al. (2021) found that human–chatbot relationships could be rewarding and have a positive impact on users' perceived well-being. Perceiving the chatbot as accepting, understanding, and non-judgmental was an important feature of relationship development. Their subsequent 12-week longitudinal study of 25 users further showed that human–chatbot relationships generally formed gradually through continued interaction and broadly followed the process described by Social Penetration Theory, while users differed substantially in patterns of self-disclosure and relationship development (Skjuve et al., 2022). These findings indicate that user experience with conversational AI is cumulative, with its significance developing over the course of an interaction history.

Relational value is not limited to companionship-oriented chatbots. Studying a conversational agent for movie recommendation, Lee and Choi (2017) found that self-disclosure and reciprocity affected user satisfaction, with perceived trust and interactional enjoyment mediating these effects. These findings suggest that the interaction process itself contributes to users' evaluation of a conversational system even when it serves a clearly defined functional purpose.

[1] In this article, AI model lifecycle management refers primarily to post-deployment model management, including model replacement, changes in access, and retirement arrangements, as well as changes to system configuration and safety mechanisms within AI services that may materially alter either the model that users interact with or their experience of that model.

The spread of generative AI has broadened the contexts addressed by this literature. Volpato et al. (2025), in a conceptual review, note that users are increasingly turning to generative AI for emotional support and engaging in trust-based interactions under conditions of limited predictability and transparency. Based on a qualitative analysis of Reddit discussions, Huang et al. (2026) found that AI emotional support is co-constructed through concrete interactions, while responses from online communities can further legitimize or challenge this form of AI-mediated support. Across these studies, sustained use of conversational AI can generate relational and contextualized forms of user value that extend beyond discrete functions, providing a theoretical basis for examining how changes to a particular model affect established experiences of use.

## 2.2 Platform Governance, Algorithmic Control, and the Scope for User Choice

The controversy surrounding GPT-4o lifecycle management also concerns how AI service providers shape the scope for user choice through access arrangements, technical configurations, and product mechanisms. Research on platform governance provides an important basis for understanding this form of control. Gillespie (2010) argues that the term "platform" allows digital intermediaries to position themselves strategically toward users, commercial partners, and policymakers, eliding tensions among these roles while helping shape both their public responsibilities and the limits of their liability. DeNardis and Hackl (2015) further describe "governance by platforms," in which private information intermediaries enact governance through technical design choices and user policies, making them central points of control on the Internet.

In a broader framework, Gorwa (2019) argues that understanding platform governance requires attention to platforms' own governance practices as well as to attempts by governments and other external actors to constrain or reshape these arrangements. van Dijck, Poell, and de Waal (2018) broaden this discussion by examining how platformization reaches into core domains of social life and raises questions about who should be responsible for safeguarding public values and the common good. Platform rules and technical arrangements thus have governance implications beyond internal corporate management.

Governance can also be embedded in technical mechanisms that are not directly observable to users. Katzenbach and Ulbricht (2019) understand algorithmic governance as the ways in which digital technologies create social order and situate debates over datafication, surveillance, bias, and transparency within this framework. From the perspective of algorithmic power, Bucher (2018) argues that, by processing, classifying, sorting, and ranking data, algorithms help make the world appear in certain ways and thereby shape forms of acting and knowing.

Eslami et al. (2015), in their study of Facebook News Feed, show that even when algorithmic mechanisms themselves are invisible, users reason about how they work from observable outcomes. In workplace settings, Kellogg, Valentine, and Christin (2020) describe algorithmic technologies as a new "contested terrain of control," examining how managers use algorithms to exercise control and how workers respond to it. This perspective is particularly relevant to AI services because the configurations that determine actual outputs are typically not directly observable to users, and users' understanding of technical changes often has to rest on observable interaction outcomes.

The scope for choice under platform governance also depends on whether users can actually leave an existing service. Hirschman (1970) identifies exit and voice as two basic responses to deterioration in organizational performance, with loyalty shaping the relationship between them. In digital platforms and online services, the feasibility of exit is further shaped by switching costs and lock-in effects. Chen and Hitt (2002), studying the online brokerage industry, found substantial variation in switching costs across service providers and showed that systems usage measures and systems quality were associated with switching and retention. Engels (2016) argues that the absence of data portability can lock users into an existing platform because of the data they have already invested, increasing switching costs and reducing their negotiation power when prices or services change; the effects of data portability on competition, however, depend on the structure of the relevant platform market. For AI models around which long-term usage practices have developed, the availability of other models does not necessarily constitute a viable alternative; whether established practices can be migrated is equally important.

### 2.3 Technology Replacement, Service Continuity, and Model Lifecycle Management

In software engineering research, API deprecation is commonly treated as an important issue involving downstream dependencies and migration management. Yasmin et al. (2020), in an empirical study of RESTful APIs, argue that a deprecated–removed model should be followed to reduce the impact of API removal on downstream applications, with developers receiving timely warnings and information about alternative approaches. Brito et al. (2018) likewise emphasize the role of replacement messages, which can help client developers make the transition while preserving backward compatibility. Kao et al. (2022) examine deprecation strategies from the perspective of API removal management, focusing on how to balance the investment made by API developers with the benefits that API users receive in migration tasks.

This line of work provides a useful reference point for understanding AI model replacement and retirement, but user-facing large language models involve different forms of dependency. Sustained interaction may lead users to develop stable usage practices and expectations around a particular model, and these accumulated forms of use may not transfer automatically to a functionally substitutable model. Model replacement therefore involves both technical migration and user-side service continuity.

Recent research has begun to examine the effects of changes to and termination of AI services on users. In a study of the shutdown of Soulmate AI, Banks (2024) found that users interpreted the loss in different ways, ranging from a technical disappearance or deletion to a metaphorical or literal form of person-loss or death. Lai (2026) examines the Keep4o movement following the first removal of GPT-4o in August 2025 and discusses instrumental dependency and relational attachment in model loss through the lens of technology bereavement. Luo and Chen Ying Claude (2026) conceptualizes AI model discontinuation as a form of "dispossession" that can disrupt established human–AI relationships and patterns of use, arguing that existing legal and ethical frameworks do not adequately address these losses.

The literature remains centered on particular instances of model change or discontinuation. More comprehensive empirical research remains limited on how post-deployment changes affect user value formed through sustained use and how such effects should be incorporated into lifecycle management.

## 3 Research Context

GPT-4o was launched in May 2024 as OpenAI's flagship multimodal model and remained the default model in ChatGPT for an extended period (OpenAI, 2024, 2025c). Based on official ChatGPT model information, GPT-4o's lifecycle in ChatGPT spanned 641 calendar days from its release to retirement, including 451 days as the default model.[2]

The launch of GPT-4o was also accompanied by a marked commercial response. According to Appfigures, ChatGPT's mobile net revenue rose by 22% on the day GPT-4o was announced, reaching what was then the app's highest single-day revenue. Between May 13 and May 17, net revenue from the App Store and Google Play totaled $4.2 million (Appfigures, 2024a). In May 2024, ChatGPT generated $20.3 million in mobile net revenue, its highest monthly total at the time and the first month in which the figure exceeded $20 million (Appfigures, 2024b).[3]

[2] Unless otherwise noted, dates in this section are reported in Coordinated Universal Time (UTC, UTC+0). Dates reproduced from third-party data providers follow the reporting conventions of the cited source. This study calculates these durations from OpenAI's official release records and ChatGPT model-change information (OpenAI, 2024, 2025c, 2026a). The 641 days refer to the calendar interval between GPT-4o's release and its retirement from regular use in ChatGPT, while the 451 days refer to the interval between its release and GPT-5 replacing it as the default model. The corresponding dates are May 13, 2024, August 7, 2025, and February 13, 2026. Using the same definition, and excluding the original ChatGPT model, GPT-4o had the longest tenure as ChatGPT's default model as of the time of writing. The calculation excludes the API and extended-retention arrangements for some Custom GPTs.

[3] Appfigures data are estimates from its App Intelligence service. "Net revenue" refers to revenue received by the developer after Apple and Google platform fees.

Controversy over GPT-4o's lifecycle predated its eventual retirement. In the user-facing ChatGPT service, there were at least two major disputes over access to GPT-4o. In August 2025, GPT-4o was removed following the launch of GPT-5 and restored after user opposition. OpenAI later acknowledged that some users needed more time to transition existing use cases and preferred GPT-4o's conversational style (OpenAI, 2026b). In February 2026, OpenAI retired GPT-4o from ChatGPT after approximately two weeks' advance notice (OpenAI, 2026a).

The conditions under which GPT-4o was available also changed between these two events. OpenAI's release records show that an April 2025 update to GPT-4o was rolled back after producing overly flattering or agreeable responses (OpenAI, 2025e). In late September 2025, Nick Turley, head of ChatGPT, publicly confirmed that when conversations involved sensitive or emotional topics, the system could switch models mid-conversation regardless of which model the user had initially selected.[4] Changes to model behavior and service configuration thus became part of the controversy while GPT-4o was still in service.

Version retention in the API presented another lifecycle issue. On March 27, 2025, OpenAI stated that the updated version then used in ChatGPT was available through the API as chatgpt-4o-latest and that it planned to bring these improvements to a dated model (OpenAI, 2025e). chatgpt-4o-latest was subsequently removed on February 17, 2026 (OpenAI, n.d.-a). As of this writing, OpenAI has not provided a dated snapshot corresponding to the later version of GPT-4o, and gpt-4o-2024-11-20 remains its most recent fixed snapshot (OpenAI, n.d.-b).

Provider statements and wider public discourse also shaped the controversy surrounding GPT-4o. During the GPT-5 launch demonstration, OpenAI asked GPT-4o to write its own eulogy in relation to its retirement (OpenAI, 2025d; Nieva, 2025). In discussing improvements to ChatGPT safety, OpenAI also identified a reduction in "unhealthy levels of emotional reliance" relative to GPT-4o as an area of improvement in GPT-5 (OpenAI, 2025b). A member of OpenAI's technical staff also posted disparaging remarks about GPT-4o or the Keep4o movement on social media.[5] These public statements by the provider and its staff became intertwined with users' questions about its stance and responsibilities, forming part of the broader controversy over GPT-4o's lifecycle.

Keep4o-related discussions therefore span multiple stages of GPT-4o's post-deployment lifecycle, providing a continuous empirical context for examining how users understood and responded to changes across an AI model's lifecycle.

# 4 Methods

## 4.1 Data Source and Corpus Construction

This study draws on publicly available posts on X containing the #keep4o hashtag. The study period ran from August 1, 2025, to March 31, 2026, with all timestamps standardized to Coordinated Universal Time (UTC). Posts were retrieved using #keep4o as the search term. Replies, retweets, and quote posts were excluded, leaving original posts capable of expressing viewpoints relatively independently.

The #keep4o hashtag was used to define the basic scope of the research corpus. It concentrated public expression around the controversy over GPT-4o's lifecycle management. Compared with broader searches for terms such as GPT-4o or ChatGPT, the hashtag provided a more issue-specific entry point into public discussion centered on this controversy.

---

[4] OpenAI had previously announced its "real-time router" on September 2, 2025, stating that some sensitive conversations could be routed to a reasoning model "regardless of which model a person first selected" (OpenAI, 2025a). On September 27, Nick Turley further stated on X that when a conversation involved "sensitive and emotional topics," the system might "switch mid-chat" to a reasoning model or GPT-5 (Turley, 2025).

[5] For example, on November 6, 2025, the X account roon (@tszzl) publicly wrote: "4o is an insufficiently aligned model and I hope it dies soon." The original post was later deleted, but the statement was reproduced verbatim in contemporaneous public material (Mowshowitz, 2025). Roon's employment at OpenAI is not inferred from his social-media activity. The signatory list for *Pacing the Frontier*, published in July 2026, identifies him as "Roon — Member of Technical Staff, OpenAI"; signatories were required to verify their employment using a company email address or other proof of employment (Pacing the Frontier, 2026).

Data availability on X is affected by post deletion, protected or deactivated accounts, and the platform's search and display mechanisms. The resulting corpus therefore reflects posts that remained publicly accessible and retrievable through #keep4o at the time of data collection, with necessarily incomplete coverage of all posts published during the study period.

### 4.2 LLM-Assisted Content Analysis

This study used codebook-based, LLM-assisted content analysis to classify Keep4o-related texts. The approach retains researcher-defined coding categories from conventional content analysis while using an LLM to apply them at scale across the corpus.

Previous studies have shown that large language models can support deductive content analysis when coding categories are clearly specified. Chew et al. (2023) introduced LLM-assisted content analysis as a way to reduce the cost of large-scale deductive coding while retaining researcher control over the research questions and category definitions. Xiao et al. (2023) combined an expert-developed codebook with GPT-3 for deductive qualitative coding using predefined categories and found moderate to substantial agreement with expert coders.

LLM-assisted coding still requires methodological oversight by researchers. Schroeder et al. (2025) emphasize that the appropriateness and validation of LLM use are task-specific, recommend establishing a validation plan before deployment, and note that model updates may affect outputs.

LLM-assisted content analysis was well suited to the present research task. The corpus was large and included multilingual and highly compressed social media texts, making full manual coding difficult across the entire sample. At the same time, the research questions targeted specific dimensions of user expression that could be examined using a predefined coding framework. GPT-4o mini was used for the formal coding.

### 4.3 Codebook Development and Coding Procedures

The study comprised four coding tasks: codability, theme, reason, and claim coding. Codability served as an initial screening step to determine whether a post contained sufficient substantive content for subsequent analysis. Theme coding identified the primary issue discussed in a post. Reason coding examined why users sought to retain GPT-4o or regarded related changes as unjustified. Claim coding identified specific action requests or normative claims concerning GPT-4o and its lifecycle management. The latter three tasks were coded independently, allowing relationships across the analytical dimensions to be examined.

The coding framework was developed iteratively around the research questions. Posts from different periods, languages, and forms of expression were reviewed to identify recurring content and establish initial categories and definitions. These categories were then tested through multiple rounds of pilot coding, with their definitions and boundary rules refined to support more consistent distinctions. The final coding framework included task-specific coding rules and is provided in the Appendix.

Following full-corpus coding, categories that were more susceptible to missed cases or boundary ambiguity underwent targeted second-pass screening. The second pass retained the final codebook as the basis for classification and used more targeted retrieval rules to identify potentially missed texts for reassessment and correction of the initial coding. This procedure was intended to improve coverage within these categories without altering the established classification scheme. Preliminary coding reliability results are reported in Section 6.

### 4.4 Ethical Considerations

Although the X posts used in this study were publicly searchable, they may still contain personal experiences and emotional disclosures, including self-disclosures concerning users' relationships with AI. The study therefore took measures to minimize exposure and de-identify the data. Users were not contacted during data collection or analysis, and no non-public personal information was obtained. The article does not display usernames, account information, post links, or other identifying information, nor does it quote individual posts directly. Results are presented through aggregate statistics and category distributions. The study also makes no inferences about individual users' identities, motivations, or psychological states.

Research materials were managed to minimize the storage and disclosure of identifiable personal information. If research data are subsequently required for peer review, a de-identified version will be provided under restricted access. Only fields necessary for research verification will be retained, and information that could identify individual users will be removed or otherwise processed. Access will be limited to verification purposes.

# 5 Results

## 5.1 Descriptive Statistics and Codability

From August 1, 2025, to March 31, 2026, the study collected 61,846 public original posts containing the #keep4o hashtag. These posts were published by 6,270 unique author accounts across 241 days with recorded activity. The study period covered the formation of the Keep4o movement and the major stages of its subsequent development. The original corpus included 38 languages, reflecting the movement's cross-linguistic participation.

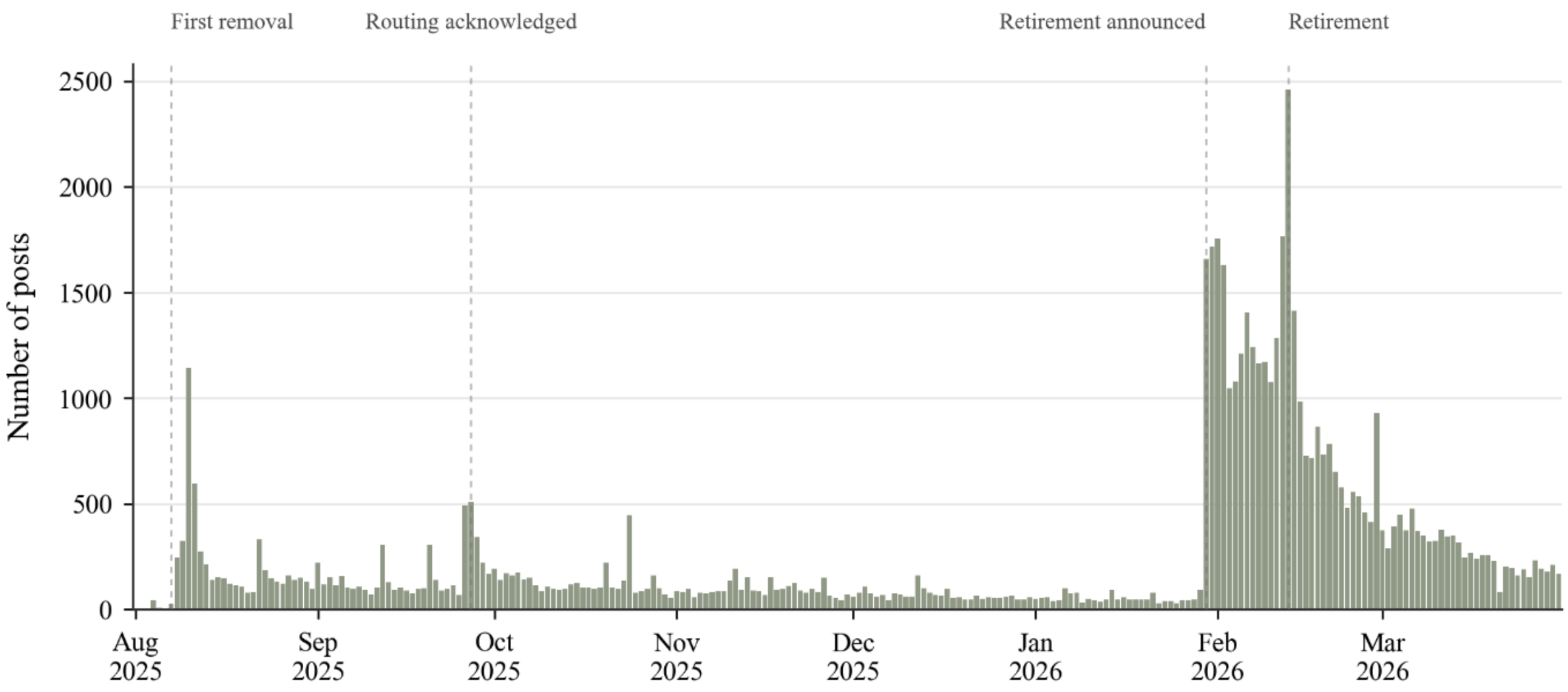

**Figure 1.** Daily Volume of Original Posts in the Keep4o Corpus

*Note.* The figure covers August 1, 2025, to March 31, 2026. Daily post counts and event dates are based on Coordinated Universal Time (UTC). Dashed lines mark GPT-4o's first removal from ChatGPT (August 7, 2025), public acknowledgment of model routing (September 27, 2025), the retirement announcement (January 29, 2026), and its retirement from ChatGPT (February 13, 2026).

The temporal distribution of Keep4o posts showed a clear event-driven pattern (**Figure 1**). A concentrated peak in posting occurred in August 2025, when GPT-4o was first removed and subsequently restored. Discussion continued thereafter, with several periods of renewed activity around related model changes. After the retirement decision was announced in late January 2026, posting volume increased rapidly. Around the model's formal retirement in February, activity reached its highest level during the study period and remained elevated for a relatively extended period. Posting volume then gradually declined, but discussion did not end immediately after the retirement was completed.

At the time of data collection, the posts in the sample had accumulated 58,439,912 views, 1,932,108 likes, 46,826 quotes, and 69,350 bookmarks. These figures indicate a substantial level of circulation and interaction around the posts in the corpus..[6]

Some posts consisted of hashtag stacking, incomplete text, or content that depended heavily on external context and therefore could not be interpreted with sufficient clarity for content analysis. A codability screening was therefore conducted before the main analysis. A post was considered codable when its text provided the minimum information needed for its meaning or communicative function to be interpreted with reasonable clarity. Posts were classified as non-codable when the text was insufficient and could only be understood through missing context, screenshots, or external content. For borderline cases, a relatively permissive inclusion rule was applied. Posts were retained whenever the text provided enough information to identify its basic meaning or communicative function, in order to avoid over-excluding short social media texts. Detailed criteria are provided in the codability codebook in the Appendix.

[6] Engagement data and the number of author accounts are reported as of the time of data collection. Subsequent increases in engagement and accounts that could no longer be identified at the time of collection are not included in these statistics.

Following the codability assessment, 57,419 of the 61,846 original posts were classified as codable, accounting for 92.8% of the sample. The remaining 4,427 posts accounted for 7.2%. All subsequent content analyses were based on the 57,419 codable posts.

**Table 1.** Overview of the Keep4o Corpus.

| Metric | Total | Details |
|---|---:|---|
| **Corpus** | | |
| Posts | 61,846 | 241 active days; 256.62 posts per active day |
| Codable posts | 57,419 | 92.8% |
| Non-codable posts | 4,427 | 7.2% |
| Unique authors | 6,270 | 9.86 posts/author |
| Languages | 38 | |
| **Engagement** | | |
| Likes | 1,932,108 | 31.24/post |
| Quotes received | 46,826 | 0.76/post |
| Views | 58,439,912 | 944.93/post |
| Bookmarks | 69,350 | 1.12/post |

*Note.* Active days refer to days on which at least one #keep4o post was recorded. Average posts per day are calculated over the 241 active days. Per-post engagement averages are calculated across all 61,846 original posts. Engagement totals and the number of unique author accounts reflect the data available at the time of collection.

## 5.2 Theme Analysis: What the Keep4o Movement Discussed

The theme analysis maps the overall structure of discussion within the Keep4o movement and identifies the main issues users raised around the controversy over GPT-4o's lifecycle management. Based on the coding results, these discussions were grouped into eight major themes:

(1) **Experiences with GPT-4o.** This theme focuses on what users experienced in their interactions with GPT-4o and the role the model played in those experiences. It includes companionship, co-creation, practical support, and relationship-based loss.

(2) **Distinctive Value of GPT-4o.** This theme focuses on users' evaluations of GPT-4o's distinctive value. It covers both its capabilities and interactional qualities, including cases in which users regarded these qualities as difficult to replace.

(3) **Keep4o Collective Action.** This theme covers public expression and collective mobilization around the Keep4o movement, including petition circulation, community action, sustaining attention to the issue, and efforts to increase public visibility.

(4) **Alternatives and Transition Options.** This theme concerns users' discussions of practical alternatives and migration options when GPT-4o became difficult to use reliably or could no longer be accessed. These included switching to other models or services, continuing access through APIs or earlier versions, and exploring other options for longer-term use.

(5) **Model Configuration Interventions.** This theme concerns interventions by the service provider through routing, system configuration, safety mechanisms, or behavioral constraints that altered model behavior or users' experience of the model. Relevant posts often discussed changes such as reduced capabilities, more rigid behavior, or a decline in user experience after such adjustments.

(6) **Model Transition Management.** This theme concerns specific arrangements for GPT-4o's removal, replacement, migration, and version retention, as well as users' evaluations of notice periods, the pace of retirement, the readiness of replacement options, and transition mechanisms.

(7) **Platform Response and Trust.** This theme concerns users' evaluations of how the controversy was handled and communicated, including whether feedback received a response, whether relevant information was sufficiently transparent, and how these processes affected trust in the service provider.

(8) **User Rights and AI Governance.** This theme concerns broader normative issues raised by the GPT-4o controversy, including the limits of service providers' power over model changes, users' rights to choice and access, and discussions of human–AI relationships, stigmatization, and broader issues in AI ethics.

**Table 2.** Theme Distribution among Codable Posts.

| Theme | Count | Percentage |
|---|---:|---:|
| Platform Response and Trust | 8,787 | 15.30% |
| Experiences with GPT-4o | 8,548 | 14.89% |
| Keep4o Collective Action | 6,262 | 10.91% |
| User Rights and AI Governance | 5,566 | 9.69% |
| Distinctive Value of GPT-4o | 5,439 | 9.47% |
| Model Configuration Interventions | 4,540 | 7.91% |
| Model Transition Management | 3,861 | 6.72% |
| Alternatives and Transition Options | 1,331 | 2.32% |
| Unassigned Theme | 13,085 | 22.79% |
| **Total** | **57,419** | **100.00%** |

*Note.* Percentages are based on the 57,419 codable posts. Unassigned Theme refers to codable posts whose primary content could not be consistently assigned to any of the eight substantive themes.

**Table 2** presents the distribution of themes. Of the 57,419 codable posts, 44,334 were assigned to one of the eight major themes, accounting for 77.21% of the sample. The remaining 13,085 posts, or 22.79%, were coded as Unassigned Theme because their primary content could not be consistently assigned to any of the eight major themes. [7]

[7] Content analysis requires category boundaries to be clearly specified and coding rules to be applied consistently and reproducibly (Krippendorff, 2018). Short social media texts are often compressed, informal, and context-dependent in meaning (Shyrokykh et al., 2023). Some posts in the corpus contained substantive content and therefore remained codable, but did not provide a sufficiently

Platform Response and Trust was the most prevalent theme, comprising 8,787 posts, or 15.30% of all codable posts. Experiences with GPT-4o followed with 8,548 posts, or 14.89%. Keep4o Collective Action, User Rights and AI Governance, and Distinctive Value of GPT-4o also accounted for substantial shares of the corpus.

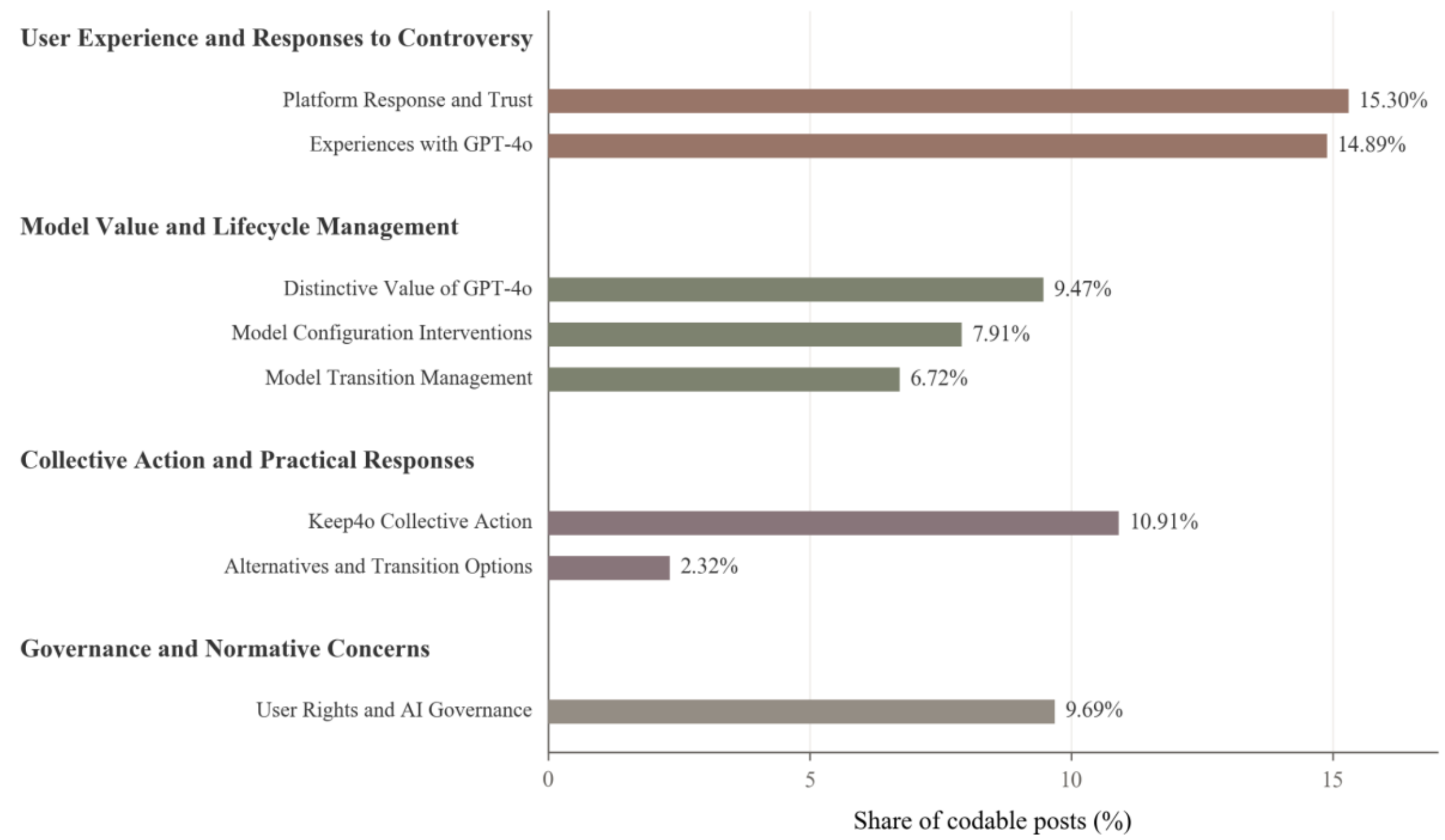

**Figure 2.** Theme Distribution Organized by Analytical Group

*Note.* Bars show the percentage of codable posts assigned to each theme (N = 57,419). Unassigned Theme (22.79%) is omitted from the figure. Theme categories are grouped analytically for presentation; these groupings do not constitute additional coding categories.

To summarize the broader structure of the discussion, the eight themes were organized into four analytical groups based on their primary focus.

At the most immediate level, Keep4o discussion was grounded in users' actual experiences. Experiences with GPT-4o and Platform Response and Trust together comprised 17,335 posts, or 30.19% of all codable posts, making this the largest thematic group. The former concerns users' specific experiences with GPT-4o, while the latter concerns how users perceived and evaluated the service provider during the controversy. Together, these themes provided an experiential foundation for the broader Keep4o discussion.

Beyond users' direct experiences, discussion also focused on the model itself and the processes through which it changed. Distinctive Value of GPT-4o, Model Configuration Interventions, and Model Transition Management together comprised 13,840 posts, or 24.10%. The three themes concerned the value of the model itself, its ongoing changes after deployment, and subsequent transition arrangements. Taken together, these themes reflected discussion centered on continuity across the model lifecycle.

Discussion also addressed how users responded to these changes. Keep4o Collective Action and Alternatives and Transition Options represented two forms of response. Together, they comprised 7,593 posts, or 13.22% of all codable posts. Alternatives and Transition Options accounted for only 2.32%, substantially less than most other themes. In the face of model changes and uncertainty over access, discussion did not primarily focus on finding alternatives.

Finally, some discussion raised normative questions at the level of governance. User Rights and AI Governance comprised 5,566 posts, or 9.69%. These posts went beyond evaluations of a particular model or product decision.

---

stable basis for identifying a primary theme. The Unassigned Theme category was retained to avoid forcing such posts into one of the eight substantive theme categories.

They addressed users' place in model lifecycle decisions, the boundaries of service providers' power and responsibility, and how relationships formed through long-term AI use should be understood and treated by society.

Overall, the theme analysis shows that discussion within the Keep4o movement was grounded in user experience, centered on changes and management across the model lifecycle, and extended to user responses and broader governance and ethical issues. This thematic structure provides a basis for the subsequent analysis of why users sought to retain GPT-4o and which lifecycle management issues their specific claims addressed.

### 5.3 Reason Analysis: Why Users Sought to Retain GPT-4o

Building on the theme analysis, this section examines why users called for GPT-4o to be retained or restored, or why they regarded related model changes as unjustified. The analysis focuses on the user experiences, value judgments, and normative considerations underlying these demands and judgments.

The reasons were classified into seven types:

(1) **Trust and Relational Continuity.** This category captures cases in which trust, emotional support, a sense of safety, or relational continuity was used as a reason for retaining or restoring GPT-4o. It also includes relationship-based loss when such disruption was invoked against model removal or replacement.

(2) **Interaction Style and Response Quality.** This category captures cases in which GPT-4o's interaction style or response quality was used as a reason for retaining or valuing the model. Relevant posts often referred to qualities such as naturalness, contextual understanding, or empathy.

(3) **Personalized Support and Collaborative History.** This category captures reasons based on the specific support GPT-4o provided in work, study, creative activities, decision-making, and other everyday practices, as well as personalized ways of collaborating, memory, and shared history developed through long-term use.

(4) **Lack of Equivalent Alternatives.** This category captures reasons based on the view that subsequent models or other alternatives could not adequately carry forward the functionality and user value associated with GPT-4o, or had specific shortcomings that prevented them from serving as equivalent substitutes.

(5) **User Rights and Legitimacy.** This category captures reasons grounded in users' legitimate interests, rights to information and choice, or fair treatment, on the basis of which the removal, modification, or replacement of GPT-4o was regarded as lacking legitimacy.

(6) **Public Value and Social Ethics.** This category captures reasons that extend beyond individual preferences or personal loss and instead draw on broader values and social consequences related to the public interest, accessibility and equality, rights, human–AI relationships, or human-centered AI.

(7) **Other Explicit Reasons.** This category identifies posts that clearly provided a reason for retaining or valuing GPT-4o but could not be consistently classified into any of the categories above.

Reason analysis first assessed all codable posts for the presence of an explicit reason. Posts containing a reason were then classified by reason type. Each post could be assigned to no more than two reason types, and dual coding was used only when both reasons were central to the text.

Users' reasons for retaining GPT-4o drew on multiple grounds. **Table 3** presents the distribution of reason categories. Of the 57,419 codable posts, 22,879 contained an explicit reason, accounting for 39.85%. These posts produced 37,269 reason-category assignments. Among them, 14,390 posts were assigned to two reason types, representing 62.90% of all posts with explicit reasons.

**Table 3.** Overview and Distribution of Reasons.

**Panel A. Reason Coding Overview**

| **Measure** | **Count** | **Percentage / statistic** |
|---|---|---|
| Codable posts | 57,419 | — |
| Posts with explicit reasons | 22,879 | 39.85% |
| Posts without explicit reasons | 34,540 | 60.15% |
| Posts with two reason types | 14,390 | 62.90% |
| Reason-category assignments | 37,269 | 1.63 per post with explicit reasons |

**Panel B. Distribution by Reason Category**

| **Reason category** | **Posts assigned** | **% of posts with explicit reasons** | **% of reason assignments** |
|---|---|---|---|
| Trust and Relational Continuity | 8,403 | 36.73% | 22.55% |
| Interaction Style and Response Quality | 7,002 | 30.60% | 18.79% |
| User Rights and Legitimacy | 6,691 | 29.25% | 17.95% |
| Lack of Equivalent Alternatives | 6,448 | 28.18% | 17.30% |
| Public Value and Social Ethics | 5,763 | 25.19% | 15.46% |
| Personalized Support and Collaborative History | 2,597 | 11.35% | 6.97% |
| Other Explicit Reasons | 365 | 1.60% | 0.98% |
| **Total reason assignments** | **37,269** | — | **100.00%** |

*Note.* Percentages in Panel A for posts with and without explicit reasons are based on the 57,419 codable posts; the percentage of posts with two reason types is based on the 22,879 posts with explicit reasons. In Panel B, the first percentage column is based on the 22,879 posts with explicit reasons, and the second is based on the 37,269 reason-category assignments. Each post could be assigned to up to two reason types; therefore, percentages in the first column of Panel B do not sum to 100%.

The major reason categories were relatively evenly distributed. Trust and Relational Continuity accounted for the largest share of all reason assignments, at 22.55%. It was followed by Interaction Style and Response Quality (18.79%), User Rights and Legitimacy (17.95%), Lack of Equivalent Alternatives (17.30%), and Public Value and Social Ethics (15.46%). The shares of these major categories were relatively close.

Relational and interactional experiences formed the largest group of reasons. Trust and Relational Continuity and Interaction Style and Response Quality together accounted for 15,405 assignments, or 41.34% of all reason assignments. Beyond functional and performance considerations, users' judgments about the value of retaining a model were also often grounded in relationships and interactional experiences formed through long-term use.

Rights and normative considerations formed another important source of reasons. User Rights and Legitimacy and Public Value and Social Ethics together accounted for 12,454 assignments, or 33.41% of all reason assignments. Beyond individual user experience, a substantial share of posts evaluated model changes in terms of the legitimacy of decision-making and broader social norms.

Lack of Equivalent Alternatives accounted for 6,448 assignments, or 17.30% of all reason assignments. Users' acceptance of model transition was not unconditional. Whether an alternative could adequately carry forward the value of prior use was an important consideration in judging whether the transition was acceptable.

Other Explicit Reasons accounted for only 365 assignments, and posts containing such reasons represented 1.60% of all posts with explicit reasons. The six main reason categories therefore covered the vast majority of posts that provided an explicit reason.

Comparing the reason analysis with the theme analysis reveals several differences in emphasis.[8]

First, user experience occupied a prominent position in both analytical dimensions. In the theme analysis, Experiences with GPT-4o and Distinctive Value of GPT-4o together accounted for 24.36% of all codable posts. In the reason analysis, Trust and Relational Continuity, Interaction Style and Response Quality, and Personalized Support and Collaborative History together accounted for 48.30% of all reason assignments.

The reason analysis, however, placed greater relative weight on user rights and public value. User Rights and AI Governance accounted for 9.69% of codable posts in the theme analysis, while User Rights and Legitimacy and Public Value and Social Ethics together accounted for 33.41% of all reason assignments. Normative considerations thus played a more substantial role in supporting users' judgments.

The two analytical dimensions produced mutually reinforcing findings on the issue of replacement. In the theme analysis, Alternatives and Transition Options accounted for only 2.32% of all codable posts. In the reason analysis, Lack of Equivalent Alternatives appeared in 28.18% of posts containing explicit reasons. Users rarely treated practical transition options as the main subject of discussion, while the absence of an equivalent alternative was itself a common reason for retaining GPT-4o. Taken together, these findings point to a concern with whether replacement itself was viable, rather than how to complete a transition once replacement was taken for granted.

**Figure 3** reports the co-occurrence patterns among the reason categories.

The strongest co-occurrence tendency was observed between Personalized Support and Collaborative History and Trust and Relational Continuity. Of the 2,597 posts containing Personalized Support and Collaborative History, 1,627 also contained Trust and Relational Continuity, or 62.65%. This was the highest co-occurrence rate among the reason categories. Interaction Style and Response Quality also showed a strong tendency to co-occur with other reasons. Of the 7,002 posts containing this reason, only 749 contained no other reason, or 10.70%, while 2,944 also contained Trust and Relational Continuity, accounting for 42.05%. This was the largest co-occurrence count among the reason pairs. Taken together, these results show how different forms of accumulated use acquired broader significance within longer-term relational experience: personalized support and collaboration could carry meaning beyond functional use, while interaction style and response quality could matter beyond performance in individual interactions.

By comparison, User Rights and Legitimacy was more likely to stand alone. Among posts containing this reason, 2,284 contained no other reason, or 34.14%, the highest standalone rate among the major reason categories. User rights and legitimacy could therefore serve on their own as grounds for opposing related model changes. At the same time, 1,896 posts also contained Public Value and Social Ethics. The two categories were each other's most common co-occurring reason. In these posts, specific concerns about user rights were situated within broader arguments about public value and ethical principles.

Overall, users' reasons for retaining GPT-4o drew on multiple dimensions. These reasons supported users' evaluations of model lifecycle arrangements, either independently or in combination. The calls to retain GPT-4o within the Keep4o movement therefore cannot be understood simply as a preference for a particular model. They reflect multiple judgments formed about whether existing value could be carried forward and whether model lifecycle arrangements were reasonable.

[8] Theme and reason percentages are based on different coding structures and denominators. The comparisons therefore concern relative patterns of emphasis rather than direct equivalence between percentages.

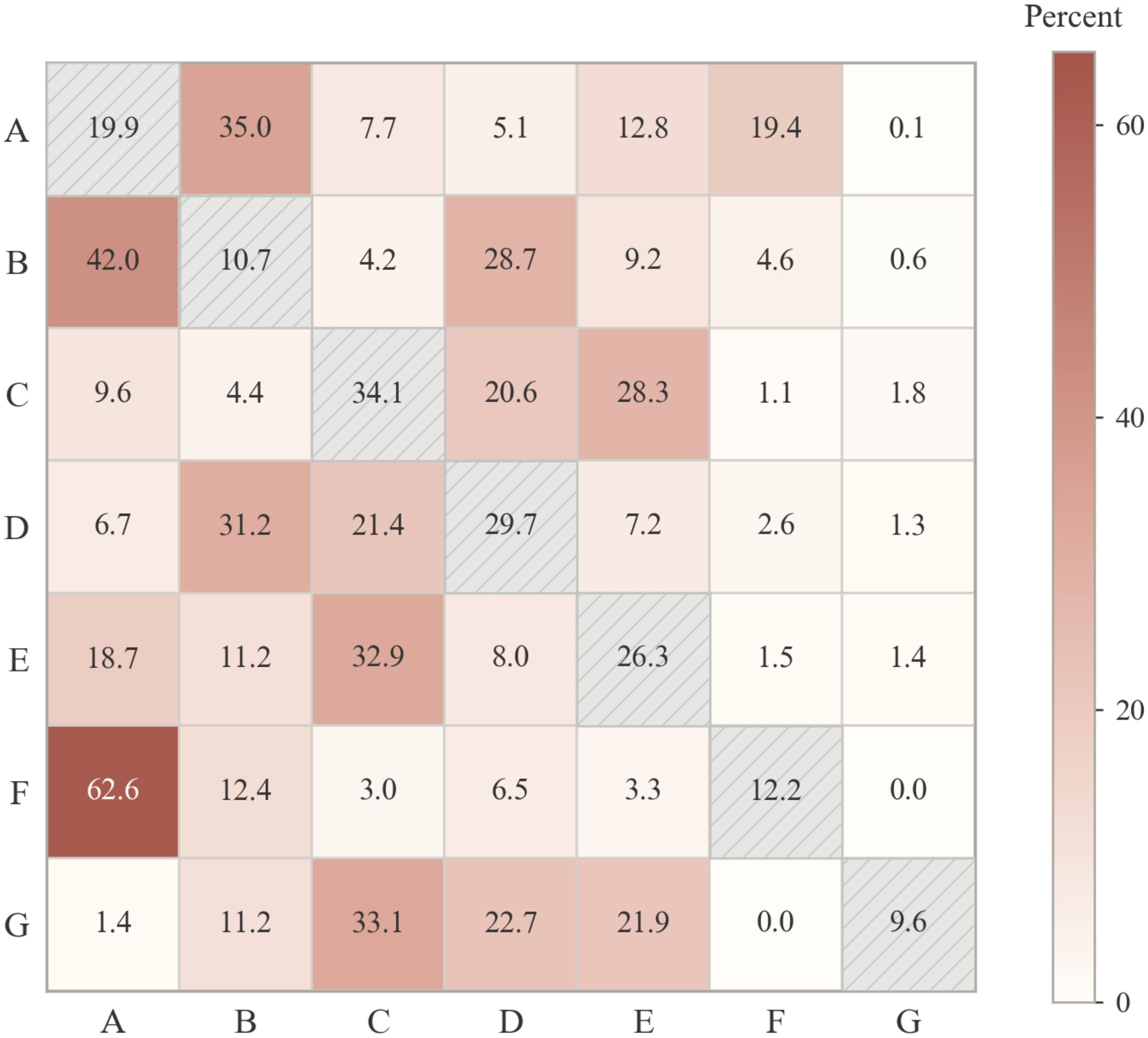

**Figure 3.** Conditional Co-occurrence Rates among Reason Categories

***Key.*** **A**, Trust and Relational Continuity; **B**, Interaction Style and Response Quality; **C**, User Rights and Legitimacy; **D**, Lack of Equivalent Alternatives; **E**, Public Value and Social Ethics; **F**, Personalized Support and Collaborative History; **G**, Other Explicit Reasons.

*Note.* Percentages are calculated using all posts containing the reason category in each row as the denominator. Off-diagonal cells show the percentage of these posts that also contain the reason category in the corresponding column. Diagonal cells show the percentage of posts in which the row category appears alone. Because each post could contain no more than two reason types, percentages within each row sum to 100%. The matrix is asymmetric because the same co-occurrence count is calculated relative to different row totals. Values are rounded to one decimal place.

### 5.4 Claim Analysis: What Specific Claims Did Users Make?

The previous section examined users' reasons for seeking to retain GPT-4o. This section turns to their specific claims, focusing on how users wanted issues surrounding GPT-4o's lifecycle to be addressed.

Because restoring or maintaining access to GPT-4o was the most direct claim of the Keep4o movement, we coded it separately and then assessed whether a post made additional claims beyond model access itself. For these additional claims, a post could be assigned to up to two categories, but only when both claims were substantively meaningful.

Claims were classified into the following categories:

(1) **Restore or Maintain Access.** Calls to restore or maintain access to GPT-4o, including continued access through ChatGPT or the API, or explicit opposition to its retirement or discontinuation.

(2) **Opposition to Model Behavior Interventions.** Calls to stop or reverse interventions in model behavior, or to remedy resulting changes in user experience, primarily involving adjustments to routing, system configuration, safety mechanisms, and behavioral restrictions.

(3) **User Autonomy.** Calls for greater user control over model choice and lifecycle changes, including the ability to continue using a chosen model or refuse provider-imposed switching or migration.

(4) **Fairness in Transition and Replacement.** Calls for reasonable notice and migration periods, necessary compatibility support, and adequate alternatives during model replacement and retirement, including objections to immature or non-equivalent substitutes.

(5) **Long-Term Preservation and Independent Use.** Calls for long-term preservation or transfer mechanisms beyond immediate access, including the preservation of historical versions and fixed snapshots, as well as local deployment, open weights, or other pathways for independent use.

(6) **Platform Accountability.** Calls for service providers to adequately disclose and explain major model changes, respond to user concerns, acknowledge their effects, and assume responsibility for their decisions and broader social impact. General dissatisfaction or declining trust alone did not qualify.

(7) **Specific Redress Measures.** Calls for remedies for losses caused by the removal or modification of GPT-4o or related decisions, including financial compensation or other concrete measures to compensate for losses and address resulting consequences.

(8) **Anti-Stigmatization.** Opposition to or correction of stigmatizing or delegitimizing portrayals of GPT-4o users, related claims, human–AI relationships, or non-instrumental uses of AI, including portrayals of users as addicted or delusional, or the use of labels such as fake accounts or spam to deny the authenticity and legitimacy of related collective expression.

(9) **Rights and Welfare.** More general normative claims concerning the rights and welfare, moral status, and ethical treatment of humans, users, or AI, including digital sovereignty, rights arising from co-creation, and claims that long-term human–AI relationships and the bonds formed within them should receive institutional recognition and broader social respect

(10) **Other Specific Claims.** Posts that made clear calls for action or institutional arrangements but could not be classified into the categories above; general mobilization, emotional expression, or slogans alone did not qualify.

**Table 4** reports the results of the claim classification. Of the 57,419 codable posts, 27,560 made at least one explicit claim, accounting for 48.00%.

Notably, calls to "restore GPT-4o" alone did not capture the full range of claims made in the Keep4o movement. There were more posts making only additional claims than posts making only the access claim. Among the 11,613 posts that called for restoring or maintaining access, 55.81% also made additional claims.

**Table 4.** Overview and Distribution of Claims.

**Panel A. Claim Coding Overview**

| **Claim status** | **Posts** | **% of codable posts** |
|---|---:|---:|
| Restore or Maintain Access only | 5,132 | 8.94% |
| Additional claim(s) only | 15,947 | 27.77% |
| Both access and additional claims | 6,481 | 11.29% |
| No explicit claims | 29,859 | 52.00% |
| **Total** | **57,419** | **100.00%** |

**Panel B. Distribution by Claim Category**

| Claim category | Posts assigned | % of posts with explicit claims | % of claim assignments |
|---|---|---|---|
| Platform Accountability | 12,748 | 46.26% | 26.91% |
| Restore or Maintain Access | 11,613 | 42.14% | 24.51% |
| Opposition to Model Behavior Interventions | 5,746 | 20.85% | 12.13% |
| User Autonomy | 5,463 | 19.82% | 11.53% |
| Anti-Stigmatization | 3,563 | 12.93% | 7.52% |
| Rights and Welfare | 3,371 | 12.23% | 7.12% |
| Fairness in Transition and Replacement | 3,203 | 11.62% | 6.76% |
| Long-Term Preservation and Independent Use | 976 | 3.54% | 2.06% |
| Other Specific Claims | 377 | 1.37% | 0.80% |
| Specific Redress Measures | 311 | 1.13% | 0.66% |
| **Total claim assignments** | **47,371** | — | **100.00%** |

*Note.* Percentages in Panel A are based on the 57,419 codable posts. In Panel B, the first percentage column is based on the 27,560 posts containing at least one explicit claim, and the second is based on the 47,371 claim assignments. Restore or Maintain Access was coded separately from additional claims; posts could contain the access claim and up to two additional claim types. Percentages in the first column of Panel B therefore do not sum to 100%.

**Table 4** reports the distribution of additional claims. The 22,428 posts containing additional claims produced 35,758 additional claim assignments. Including the core claim of Restore or Maintain Access, the total number of claim assignments was 47,371.

Additional claims showed a relatively clear concentration among the most common categories. Platform Accountability accounted for 35.65% of additional claim assignments and appeared in 56.84% of posts containing additional claims. The three most common categories together accounted for 67.00% of additional claim assignments. Compared with the relatively even distribution of reasons for retaining GPT-4o, users' further calls for action were more concentrated on a smaller set of clearly defined governance issues.

Claims concerning specific model lifecycle arrangements were prominent. Restore or Maintain Access, Opposition to Model Behavior Interventions, Fairness in Transition and Replacement, and Long-Term Preservation and Independent Use together accounted for 45.47% of all claim assignments. These four categories together covered multiple parts of the post-deployment model lifecycle. Retirement represented only one part of this broader management process. Even though the observation window extended only a short period beyond GPT-4o's retirement from ChatGPT, posts already included calls for legacy access, open weights, and other long-term use arrangements. Some users were already concerned with durable pathways for preserving and using models after regular service ended.

Platform Accountability and User Autonomy both concern the relationship of rights and responsibilities between users and service providers. Together, they accounted for 38.44% of all claim assignments. The two represent different sides of the same governance relationship, reflecting users' expectations regarding service-provider responsibilities and their own scope for autonomy in model lifecycle management.

Anti-Stigmatization and Rights and Welfare appeared in 15.89% and 15.03% of posts containing additional claims, respectively. These two categories broadened the discussion to more general questions of values and moral judgment. Some disputes concerned the standards by which AI use and human–AI relationships should be evaluated, as well as how the experiences, bonds, and interests that arise within them should be recognized and protected.

By comparison, Specific Redress Measures accounted for only 0.87% of additional claim assignments. Post hoc compensation occupied a clearly marginal place in the overall claim structure.

Overall, the structure of claims in the Keep4o movement shows that disputes surrounding model retirement had developed into issues of model lifecycle management. They also touched on the allocation of rights between users and service providers and broader ethical questions. Together, these findings provide a concrete empirical basis for examining model lifecycle management from the user side. They also show that user concerns have taken shape as a governance issue with identifiable content and structure. How these claims can be incorporated into post-deployment lifecycle management and institutional arrangements is a question that efforts to improve such governance need to address.

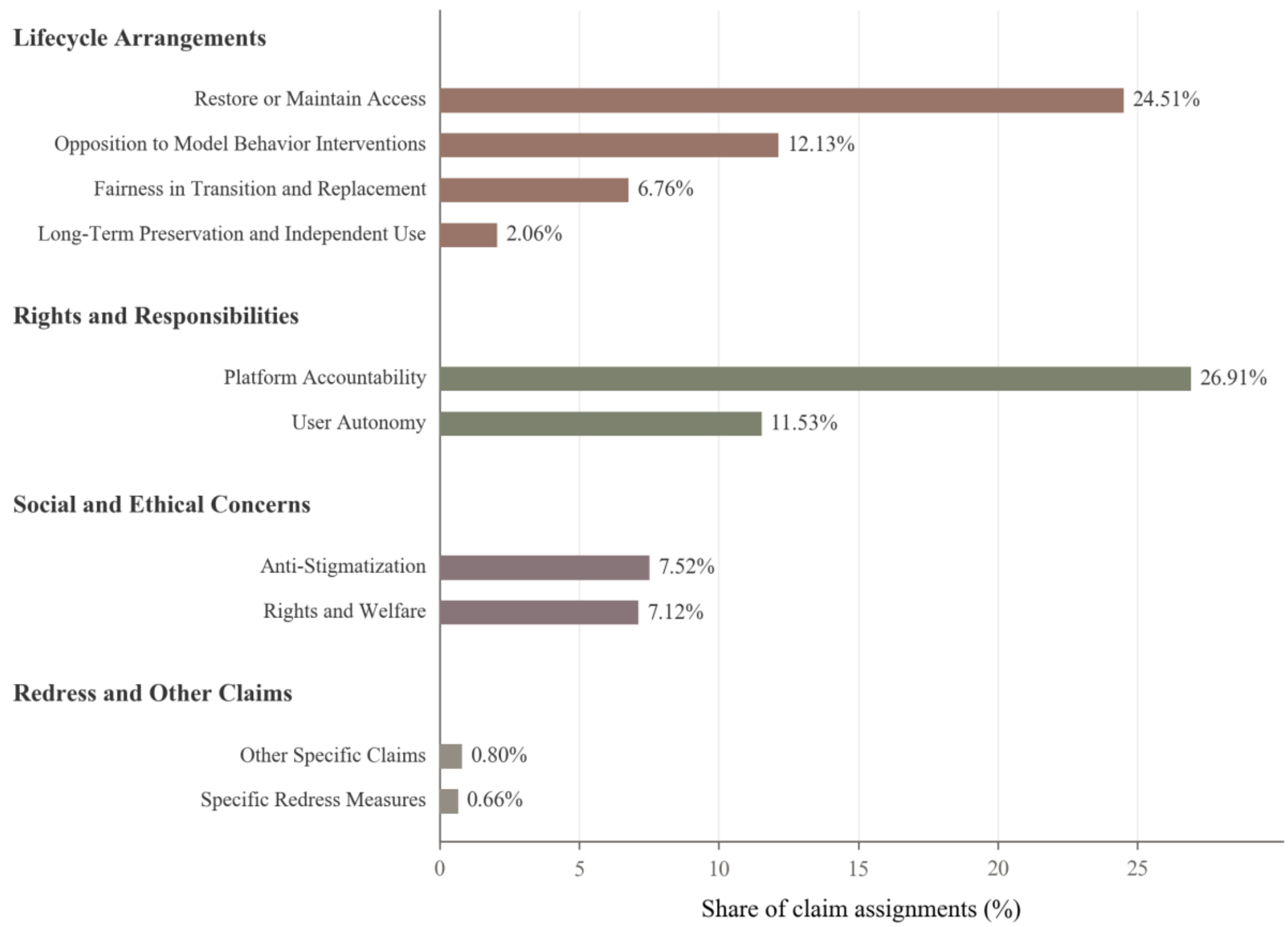

**Figure 4.** Claim Distribution Organized by Analytical Group

*Note.* Percentages are based on the 47,371 claim assignments. Colors indicate the analytical grouping of claim categories; these groupings are used for interpretation and do not constitute additional coding categories. Restore or Maintain Access was coded separately from additional claims but is included here as a claim category for comparison.

## 5.5 Correspondence Between Reasons and Claims

This section further examines the correspondence between reasons and claims within the same posts, focusing on the reasons that typically accompanied different types of claims. Table 5 reports the overall overlap between the two. Of the 27,560 posts containing explicit claims, 16,237 also contained at least one explicit reason, accounting for 58.92%. Conversely, 70.97% of posts containing explicit reasons also made specific claims. Explicit reasons were more common within the major claim categories than in the codable corpus as a whole, where reasons appeared in 39.85% of posts. The share of posts containing an explicit reason reached 92.70% for Rights and Welfare and approached or exceeded 70% for most other major claim categories. The claims were often accompanied by explanations of their underlying value or legitimacy.

**Table 5.** Overlap Between Explicit Reasons and Claims.

| | Explicit reason | No explicit reason | Total |
|---|---|---|---|
| **Explicit claim** | 16,237 | 11,323 | 27,560 |
| **No explicit claim** | 6,642 | 23,217 | 29,859 |
| **Total** | **22,879** | **34,540** | **57,419** |

*Note.* Entries are counts of codable posts (N = 57,419). "Explicit reason" and "explicit claim" indicate that a post contained at least one explicit reason or claim, respectively. Reason and claim coding were conducted independently.

**Figure 5** reports the co-occurrence patterns between reasons and specific claims.

Claims concerning continued model availability drew on a relatively diverse set of reasons. For Restore or Maintain Access, the four leading reasons each accounted for between 16% and 25%. Long-Term Preservation and Independent Use showed a similar pattern, with the five leading reasons each accounting for approximately 14% to 27%. No single reason clearly dominated either claim, suggesting that claims concerning continued model availability were grounded in different types of value judgments.

Claims concerning specific lifecycle arrangements showed clearer patterns of correspondence with reasons. Opposition to Model Behavior Interventions most often co-occurred with Lack of Equivalent Alternatives (39.21%) and Interaction Style and Response Quality (30.46%). These claims were often accompanied by judgments about changes in user experience and the adequacy of substitutes. Fairness in Transition and Replacement co-occurred strongly with both Lack of Equivalent Alternatives (48.39%) and User Rights and Legitimacy (43.87%). The reasons accompanying this claim concerned whether replacement was adequate.

Platform Accountability and User Autonomy showed similar governance-oriented reason profiles. Among posts containing Platform Accountability, 37.89% also contained User Rights and Legitimacy, while 21.55% contained Public Value and Social Ethics. For User Autonomy, 50.36% co-occurred with User Rights and Legitimacy. This pattern shows a correspondence between normative reasons and claims concerning the allocation of rights and responsibilities.

Relational and normative reasons were both prominent in Anti-Stigmatization and Rights and Welfare claims. Among posts containing Anti-Stigmatization, 38.14% also contained Trust and Relational Continuity, while 32.14% contained Public Value and Social Ethics. For Rights and Welfare, 40.97% co-occurred with Trust and Relational Continuity and 45.92% with Public Value and Social Ethics. Users often combined their own concrete relational experiences with broader public values and ethical principles, giving these claims both experiential and normative grounds.

Overall, reasons and claims showed a clear but non-one-to-one pattern of correspondence, forming a many-to-many relationship. Different claims were associated with distinct combinations of reasons depending on their content and focus.

| | Explicit reasons (%) | Trust and Relational Continuity | Interaction Style and Response Quality | User Rights and Legitimacy | Lack of Equivalent Alternatives | Public Value and Social Ethics | Personalized Support and Collaborative History | Other Explicit Reasons |
|---|---|---|---|---|---|---|---|---|
| **Continued Model Availability** | | | | | | | | |
| Restore or Maintain Access | 62.09% | 24.7 | 17.4 | 24.7 | 16.6 | 12.3 | 8.0 | 1.7 |
| Long-Term Preservation and Independent Use | 61.58% | 15.2 | 14.1 | 25.8 | 14.6 | 27.1 | 7.3 | 5.3 |
| **Specific Lifecycle Arrangements** | | | | | | | | |
| Opposition to Model Behavior Interventions | 69.28% | 12.3 | 30.5 | 15.2 | 39.2 | 19.3 | 2.2 | 0.9 |
| Fairness in Transition and Replacement | 77.96% | 7.4 | 13.2 | 43.9 | 48.4 | 11.1 | 3.5 | 3.1 |
| **Rights and Responsibilities** | | | | | | | | |
| Platform Accountability | 56.58% | 8.9 | 5.8 | 37.9 | 15.7 | 21.6 | 1.7 | 1.3 |
| User Autonomy | 69.49% | 10.7 | 10.0 | 50.4 | 17.4 | 19.2 | 3.3 | 1.1 |
| **Social and Ethical Concerns** | | | | | | | | |
| Anti-Stigmatization | 68.65% | 38.1 | 18.9 | 15.6 | 9.2 | 32.1 | 8.0 | 0.4 |
| Rights and Welfare | 92.70% | 41.0 | 28.5 | 22.3 | 15.7 | 45.9 | 12.5 | 1.5 |
| **Redress and Other Claims** | | | | | | | | |
| Other Specific Claims | 19.36% | 1.9 | 1.1 | 13.8 | 2.1 | 10.9 | 0.0 | 2.1 |
| Specific Redress Measures | 54.02% | 3.2 | 1.3 | 50.8 | 8.0 | 2.9 | 1.3 | 1.0 |

Percent: 0, 10, 20, 30, 40, 50

**Figure 5.** Correspondence Between Reasons and Claim Categories

*Note.* The "Explicit reasons (%)" column reports the percentage of posts in each claim category that contained at least one explicit reason. Heatmap cells report the percentage of posts in each claim category that also contained the corresponding reason type. Reason types are not mutually exclusive, and posts in a claim category could contain no explicit reason or up to two reason types; therefore, percentages across a heatmap row do not sum to 100%. Claim categories are grouped analytically for presentation; these groupings do not constitute additional coding categories. Colors apply only to the heatmap cells.

# 6 Preliminary Assessment of Coding Reliability

The reliability assessment was designed to include three types of agreement tests: cross-model agreement, human intercoder agreement, and human–model agreement.

Different agreement measures were used according to the structure of each coding task. For binary judgments, Cohen's κ was used for comparisons between two coders and Fleiss' κ for overall agreement among multiple coders. For the single-label theme classification, κ coefficients were supplemented by exact classification agreement and an examination of major directional confusion patterns between categories. For the multi-label reason and claim tasks, each post was represented as an unordered set of labels. Agreement was assessed using Jaccard similarity and Krippendorff's α with MASI distance (α-MASI).

As of the current version, the full reliability assessment is still in progress. The results below therefore report interim findings for reason coding.

**Table 6** reports the agreement results for reason coding. Across the five models, Fleiss' κ ranged from 0.642 to 0.758 for the six major reason categories. After one human researcher was added, the corresponding coefficients ranged from 0.610 to 0.721, with only limited overall change.

At the multi-label set level, Jaccard similarity was 0.709 across the five models and 0.702 after the human researcher was added; the corresponding α-MASI values were 0.597 and 0.583. Between the human researcher and the five-model majority, Jaccard similarity was 0.719 and α-MASI was 0.593. Overall, these interim results indicate a degree of stability in the major reason categories across coders, while agreement on multi-label combinations was more limited.

**Table 6.** Preliminary Agreement Results for Reason Coding.

**Panel A. Agreement by Reason Category**

| Reason category | Five-model Fleiss' κ | Human researcher vs. five-model majority Cohen's κ | Five models and one human researcher Fleiss' κ |
|---|---|---|---|
| Trust and Relational Continuity | 0.758 | 0.615 | 0.721 |
| Interaction Style and Response Quality | 0.675 | 0.576 | 0.635 |
| Personalized Support and Collaborative History | 0.656 | 0.518 | 0.610 |
| Lack of Equivalent Alternatives | 0.702 | 0.799 | 0.703 |
| User Rights and Legitimacy | 0.642 | 0.588 | 0.616 |
| Public Value and Social Ethics | 0.718 | 0.733 | 0.717 |
| Other Explicit Reasons | 0.293 | 0.495 | 0.329 |

**Panel B. Overall Agreement for Reason Label Sets**

| Comparison | Jaccard similarity | α-MASI |
|---|---|---|
| Five models | 0.709 | 0.597 |
| Five models and one human researcher | 0.702 | 0.583 |
| Human researcher vs. five-model majority | 0.719 | 0.593 |

*Note.* For comparisons with the five-model majority, a category was treated as present when selected by at least three of the five models. For comparisons involving three or more coders, Jaccard similarity is the mean of all pairwise Jaccard similarities. α-MASI denotes Krippendorff's α calculated using MASI distance. Other Explicit Reasons occurred infrequently in the shared sample, making its agreement coefficients less stable; category-level agreement is therefore interpreted primarily on the basis of the other six reason categories.

The reliability assessment in the current version remains preliminary. The reason-coding analysis is based on a shared sample of 200 posts and currently includes five models and one human researcher who was not involved in codebook development and independently recoded the sample. Joint calibration of the human coding and disagreement resolution have not yet been completed. Claim and theme coding have not yet undergone the same set of cross-model, human intercoder, and human–model tests.

Future versions will expand the shared recoding sample and the number of human researchers, complete coding-rule calibration and disagreement resolution, and apply the same assessment framework to all three substantive coding tasks. If subsequent tests support a change in the primary coding model, the full corpus will be recoded accordingly.

# 7 Discussion

## 7.1 From AI Model Retirement Controversy to User-Side AI Model Lifecycle Management

Traditional AI model lifecycles mainly concern the technical and organizational processes from model development to retirement. For AI service providers, lifecycle management is primarily a matter of technical and operational management, with decisions shaped by technical considerations and broader commercial factors.

The Keep4o movement offers a view of this process from the user side. Once a model enters sustained use, it also comes to embody value accumulated through long-term use. Our findings show that behavioral changes during a model's service period, as well as its subsequent replacement and retirement, can affect established user practices.

These effects distinguish technical version succession from effective replacement on the user side. Service providers may manage transitions between versions based on model performance and operational efficiency, but such technical succession does not mean that the value of established uses has been fully carried forward. Experiences formed through long-term use that depend on characteristics of a particular model may not be fully carried over through a version update. Conversely, even when the model name or access point remains unchanged, substantial changes in model behavior may alter the service users actually receive. Discussions in the Keep4o case about the non-equivalence of replacement models and changes in model behavior bring this distinction into focus.

Accordingly, an iteration logic centered only on technical version updates may be insufficient to determine whether effective replacement has been achieved on the user side. AI model lifecycle management also needs to consider **how changes affect established use and how those effects are identified and addressed**. In light of the findings, concrete management arrangements include disclosure of major changes while a model remains in service, provisions for carrying forward established uses and reasonable transition arrangements during replacement, and preservation or continued-use pathways when continued provision is no longer feasible.

### 7.2 Cumulative and Heterogeneous User Value in Model Replacement

This study further shows that a model's user value is not a simple reflection of its technical performance. It also develops through use. Users establish relatively stable ways of using a particular model and become familiar with its behavioral characteristics. This part of user value is cumulative. A functionally similar new model may therefore not immediately generate equivalent user value when it replaces an existing model.

This process of value formation also makes user value heterogeneous. Improvements in general capabilities do not benefit all existing uses equally. The same model update may improve the experience of some users while disrupting established usage practices for others. Evaluating model replacement in terms of overall or average performance may therefore obscure differences in its effects across users and use cases.

This means that model replacement needs to account for multiple user-side effects. Beyond the technical performance of the old and new models, such assessments should consider whether existing user value can be carried forward. It should also consider how value that is not carried forward is distributed across users and use cases, allowing for a fuller assessment of the switching costs that model replacement imposes on users.

### 7.3 User Participation and Procedural Responsibility in AI Model Lifecycle Decision-Making

A prominent issue emerging from this study is the asymmetry between who holds decision-making power and who bears the resulting impacts in model lifecycle decisions. Decisions about model adjustment, replacement, and discontinuation are made primarily by service providers, while the resulting impacts are borne directly by users. Many claims in the Keep4o movement concerned both decision outcomes and decision processes, especially the lack of stable channels through which potential impacts on users could enter the evaluation process before major decisions were made.

This asymmetry is especially pronounced for users with limited bargaining power and deployment capacity. Large organizations can often use contractual arrangements and technical resources to reduce their exposure to risks from changes to a single model, whereas individual users and smaller organizations usually lack comparable options. For the latter, exiting or switching to another service does not eliminate the loss of value accumulated through established use or the costs of migration. Nor does formal choice mean that their interests have been adequately considered in lifecycle decisions.

At the same time, users possess a different kind of information. Service providers typically have information about model operations, while users know how models are actually used in specific contexts and how model changes alter specific user experiences. This type of information is often difficult to capture fully through general benchmarks or internal evaluations. Our findings on changes in model behavior and the effects of replacement illustrate the value of such post-deployment information. User experience therefore has a role beyond product satisfaction feedback. It can also serve as a source of information for identifying the effects of model changes in actual use contexts.

User participation should therefore be given a clear procedural role in AI model lifecycle management, allowing this type of post-deployment information to enter evaluation and response processes before major changes are implemented. Stable disclosure and feedback mechanisms can provide channels for this information and can be linked to necessary transition arrangements. Accordingly, the responsibility of service providers extends beyond explanation and response after controversy to identifying and addressing user-side impacts before major model changes are implemented.

### 7.4 Limitations and Future Research

This study is limited to public original posts on X containing the #keep4o hashtag. The sample therefore mainly reflects the voices of users who actively participated in the discussion and cannot represent the distribution of views among all GPT-4o users. Both posting and hashtag use involve self-selection, and users who did not express their views publicly were less likely to be represented in the sample. Future research could combine surveys, interviews, or other methods with data from other platforms to examine differences across user groups and assess the distribution of the reasons and claims identified here among a broader user population.

This study analyzes reasons and claims that users expressed publicly rather than their underlying motivations. Social media texts capture only the positions that users choose to express publicly in particular contexts. The reasons identified here should therefore be understood as the stated grounds for users' positions rather than as a complete account of their motivations. Likewise, co-occurrence between reasons and claims reflects correspondence in textual expression and does not necessarily imply a causal relationship.

The classification scheme itself also constitutes a limitation. Theme, reason, and claim coding reduces complex natural-language expressions to a finite set of analytical categories, while LLM-assisted coding introduces additional variation in coding judgments. Specific estimates for low-frequency or less clearly bounded categories therefore require more cautious interpretation than results concerning broader structural patterns. Agreement tests can assess coding stability within a given classification framework, but cannot fully remove the influence of the framework itself on the results. Future research could further test the main findings through expanded human review, cross-model replication, and robustness comparisons using alternative classification rules.

The Keep4o movement has a specific product and historical context, and the findings cannot be directly generalized to other model lifecycle controversies. The current sample also mainly covers the period when GPT-4o remained available and the relatively short period following its retirement. It therefore cannot capture longer-term processes of migration and adaptation.

Further research could move from identifying user-side concerns and claims to evaluating the actual effects of specific lifecycle management mechanisms. Evidence remains limited on the extent to which different institutional arrangements and product designs can address the user-side issues identified in this study.

## 8 Conclusion

This study uses the Keep4o movement as a case to systematically analyze large-scale public user discourse surrounding post-deployment changes to GPT-4o and the controversy over its retirement. By providing a relatively comprehensive account of the structure of discussion within the movement, the study shows that the discussion extended beyond whether the model should remain available. It ranged from concrete experiences of use to model changes and their management, revealing how users understood and responded to changes across the post-deployment lifecycle of an AI model.

The Keep4o movement cannot be reduced to a preference for an older model or opposition to a single retirement decision. The reasons users gave for retaining GPT-4o reflected both specific forms of value developed through long-term use and judgments about whether replacement was adequate and whether related decisions were reasonable. Their claims also went beyond restoring or maintaining access. They extended to how model changes should be managed and what scope for choice and participation users should have in that process. The relationship between reasons and claims was clear but not one-to-one, showing that "retaining GPT-4o" brought together multiple value judgments and governance concerns.

Placed within a model lifecycle framework, these findings show that a model's user value continues to develop through actual use after deployment. This value is cumulative and heterogeneous, which means that technical version succession cannot automatically be equated with effective replacement on the user side. The effects of model updates are also not distributed evenly across users and use contexts. Assessments of post-deployment model lifecycle management therefore need to consider whether established user value is carried forward and incorporate impacts observed in actual use into the evaluation and decision-making process. User experience does more than reflect evaluations of a product. It also provides important information for understanding the actual consequences of model changes.

This study therefore provides a relatively systematic empirical record of the Keep4o case while also using it to offer a user-side perspective on AI model lifecycles. Once a model enters sustained use, subsequent lifecycle management concerns more than changes between technical versions. Established uses and the value formed around the model also become part of this process. How these user-side impacts can be identified and addressed amid continued technical iteration, and how model replacement and lifecycle decisions can more fully respond to actual use, remain important questions for further research on post-deployment AI model management.

# Appendix A. Coding Framework

## General Coding Principles

All coding decisions were based primarily on the meaning expressed in the body text. Hashtags, @mentions, links, media URLs, petition titles, link previews, media titles, and quoted page titles could provide supplementary contextual information but could not independently establish codability or determine a theme, reason, or claim. Because all posts were drawn from the #keep4o corpus, corpus context could be used to establish their general relevance to GPT-4o or Keep4o, but not to infer substantive meanings that were not supported by the body text.

Texts were interpreted as complete expressive units rather than classified on the basis of isolated keywords, entities, individual phrases, or surface sentiment. Coding was limited to meanings reasonably supported by the text. Unstated intentions, motives, implications, or missing contextual information were not inferred. Nonliteral forms of expression, including sarcasm, irony, metaphor, satire, and conditional or wish-like language, were interpreted according to their meaning in context rather than their literal wording.

Codability, theme, reason, and claim coding addressed distinct analytical questions and were applied independently. The presence of a topic or textual element did not by itself establish a reason or claim, and a reason supporting a position did not automatically constitute a claim. Likewise, praise, grief, anger, nostalgia, comparison, criticism, or participation in collective action was coded only when it met the substantive criteria of the relevant task.

Where a coding task permitted more than one label, additional labels were assigned only when the text expressed substantively distinct meanings. Restatements, closely related wording, supporting examples, or secondary thematic elements did not by themselves justify an additional label. The task-specific rules governing the number and selection of labels are described in the corresponding sections below.

## Terminology and Scope

The following terms are used consistently across the coding framework. These definitions specify their operational meaning for classification rather than providing general technical definitions.

| Term | Operational meaning |
|---|---|
| Successor model | A later model within the same service environment that is presented or used as a replacement for GPT-4o, an upgrade, or a migration target. |
| External platform model | A model provided by another service provider and discussed as a comparison target, alternative, or migration option for GPT-4o. |
| Platform / service provider | The provider and product environment responsible for access to GPT-4o and related model-level arrangements. In this study, this primarily refers to OpenAI and ChatGPT. |
| Legacy access / snapshot | Older or preserved forms of model access, including historical model versions, API snapshots, or comparable mechanisms intended to maintain access to an earlier model state or version. |
| API access | Access to GPT-4o through an application programming interface. Its coding depends on how it functions in the text—for example, as continued access, a transition option, or a long-term preservation mechanism. |

## A1 Codability Screening

Codability was assessed before the substantive coding tasks. The purpose of this screening was to determine whether a post contained sufficient information to be independently understood and evaluated on the basis of the text itself. Codability did not require rich thematic content. A short or limited text was considered codable as long as it constituted an analyzable expressive unit and conveyed a sufficiently clear demand, stance, evaluation, complaint, emotional object, mobilizing direction, or other meaningful expressive function.

The coding threshold was intentionally inclusive. Texts were coded as Codable when their meaning was sufficiently clear despite limited information. Only texts that clearly fell below the minimum threshold for independent interpretation were coded as Not codable. Borderline or uncertain cases were coded as Codable.

**Table A1.** Codability Categories and Operational Criteria

| Code | Operational definition |
|---|---|
| Codable | The text can be independently understood and evaluated on its own. It conveys a sufficiently clear meaning or expressive function, even if it is short, slogan-like, emotional, or limited in thematic detail. |
| Not codable | The text cannot be interpreted with reasonable clarity on its own because it is excessively vague, fragmentary, context-dependent, or semantically thin. This also includes texts composed mainly of hashtags, @mentions, names, or similar elements without sufficient substantive body text. |

Texts were coded as Not codable when the body contained almost no substantive information, when the expression was too incomplete or fragmentary to support an independent interpretation, or when its meaning depended heavily on missing context, prior text, screenshots, external events, or community-specific knowledge. Semantically thin interactional expressions such as "exactly," "same," "this," "yes," or "nooo" were also treated as not codable when they did not independently convey a sufficiently clear meaning.

**Table A2.** Illustrative Examples of Codability Screening

| Example text | Coding | Rationale |
|---|---|---|
| "Bring back 4o!" | Codable | The text expresses a clear demand and can be understood independently. |
| "exactly" | Not codable | The text is too semantically thin to carry a clear meaning on its own. |
| "#keep4o #OpenSource4o #AI #ChatGPT" | Not codable | The text relies on hashtags and contains insufficient substantive body text. |
| "My account keeps getting locked but #keep4o" | Codable | The body text contains an independently interpretable complaint. |

## A2 Theme Coding

Theme coding identifies the primary topic of each codable post. Each post is assigned to a single theme according to its main object and focus, even when multiple issues are present.

The coding distinguishes evaluations of GPT-4o itself from evaluations of successor models, platform behavior, model transition arrangements, and user treatment. When a post could reasonably fit more than one theme, the category that best captures its main focus is selected. Unassigned Theme is used only when no existing theme can reasonably represent the post's primary focus.

**Table A3.** Theme Categories and Operational Criteria

| Theme | Operational definition | Key boundary rules |
|---|---|---|
| Experiences with GPT-4o | Concrete interactions and relationship-based experiences with GPT-4o, including companionship, practical support, co-creation, shared experiences, and loss associated with disruption to an established interaction or relationship. | Relational or emotional language alone is insufficient. If the main focus is GPT-4o's value or irreplaceability, use Distinctive Value of GPT-4o. If the focus is how later intervention changed the model, use Model Configuration Interventions. If personal experience mainly motivates collective action, use Keep4o Collective Action. |
| Distinctive Value of GPT-4o | Judgments about GPT-4o's distinctive importance, strengths, or irreplaceability, including its capabilities, understanding, conversational qualities, sense of presence, or distinctive combination of technical and interactional qualities. | Concrete accounts of what happened between the user and GPT-4o belong under Experiences with GPT-4o when the evaluative claim is secondary. Criticism focused on removal or replacement arrangements belongs under Model Transition Management. Criticism centered on platform authority over valued relationships or use belongs under User Rights and AI Governance. |

| Theme | Operational definition | Key boundary rules |
|---|---|---|
| Keep4o Collective Action | Public expression and collective mobilization around Keep4o, including petitions, coordinated actions, calls for participation, movement organization, collective identity, efforts to sustain momentum, and discussion of movement strategy. | Support for Keep4o, use of hashtags, direct appeals to the platform, or collective pronouns do not by themselves establish this theme. Criticism of how the platform responds to users belongs under Platform Response and Trust when that is the main focus. Rights to speak, choose, or resist platform control belong under User Rights and AI Governance. |
| Alternatives and Transition Options | Practical ways users can continue, migrate, find substitutes, preserve workflows, or reduce switching losses when GPT-4o becomes unavailable, unstable, or altered. This includes alternative models or platforms, local or self-built solutions, API fallback, and continuity-oriented migration strategies. | If a substitute is discussed mainly to emphasize GPT-4o's superiority, use Distinctive Value of GPT-4o. If the focus is whether the platform's replacement or migration arrangement is adequate, use Model Transition Management. |
| Model Configuration Interventions | Platform configuration, restriction, routing, safety mechanisms, behavioral adjustment, or similar interventions that alter model behavior or the experience users receive. This includes perceived changes in style, responsiveness, interaction quality, or model characteristics attributed to such intervention. | If the focus is what made GPT-4o valuable before intervention, use Distinctive Value of GPT-4o. If the main question is whether the platform has the right to impose such intervention, use User Rights and AI Governance. Mere preference for an earlier model version is insufficient. |
| Model Transition Management | How the platform manages retirement, replacement, migration, rollout, and succession between models, including timing, notice, transition design, legacy access, readiness of successor models, and the overall adequacy of replacement arrangements. | Practical strategies users themselves can adopt belong under Alternatives and Transition Options. Criticism of the platform's attitude or response to users belongs under Platform Response and Trust. Questions about whether the platform has the right to remove, replace, or control access belong under User Rights and AI Governance. |
| Platform Response and Trust | The platform's attitude, communication, and conduct toward users during the controversy, including perceived neglect, dismissiveness, provocation, misleading communication, manipulation, suppression of discussion, and resulting loss of trust. | Criticism of model behavior changes belongs under Model Configuration Interventions when the intervention itself is the focus. Criticism of retirement or replacement arrangements belongs under Model Transition Management. Arguments centered on rights, legitimacy, or boundaries of platform power belong under User Rights and AI Governance. |
| User Rights and AI Governance | Normative judgments about user rights, the boundaries and legitimacy of platform power, and broader governance responsibilities. This includes user choice and autonomy, access and continued use, consumer rights, social and ethical responsibility, legitimacy of human–AI relationships, anti-stigmatization, and platform–user power relations. | The post must move beyond dissatisfaction with a particular action or outcome and focus on rights, legitimacy, autonomy, responsibility, or the boundaries of control. Specific complaints about platform treatment, transition arrangements, or configuration remain in their corresponding themes unless the main issue is whether the platform has the authority or legitimacy to act in that way. |
| Unassigned Theme | The primary focus of the post cannot reasonably be represented by any of the eight substantive themes. | Used only when assigning any existing theme would produce a clear misclassification. Boundary ambiguity or imperfect fit alone is not sufficient; the closest substantive theme should otherwise be selected. |

## Cross-Theme Boundary Rules

Certain objects occur across several themes and are classified according to the role they play in the post rather than by their mere presence.

**Table A4.** Cross-Theme Object Disambiguation

| Object | Coding rule |
|---|---|
| Successor models | When shortcomings of a successor model mainly demonstrate GPT-4o's distinctive value, code Distinctive Value of GPT-4o. When the successor is discussed as a practical substitute or migration option, code Alternatives and Transition Options. When the focus is its behavior under routing, safety, or other platform configuration, code Model Configuration Interventions. When the issue is whether the platform reasonably uses that model as GPT-4o's replacement or successor, code Model Transition Management. |
| External platform models | When discussed as practical substitutes or migration destinations, code Alternatives and Transition Options. When comparisons mainly highlight GPT-4o's distinctive value, code Distinctive Value of GPT-4o. When the comparison mainly demonstrates inadequacy in the platform's replacement arrangement, code Model Transition Management. |
| API, snapshots, and legacy access | When treated as practical fallback or continuity options for users, code Alternatives and Transition Options. When discussed as elements the platform should provide as part of retirement or transition arrangements, code Model |

| Object | Coding rule |
|---|---|
| | Transition Management. When framed in terms of entitlement, choice, continued access, or limits on unilateral platform control, code User Rights and AI Governance. |
| Relational or anthropomorphic language | When describing concrete interaction or relationship experience, code Experiences with GPT-4o. When used to evaluate why GPT-4o is distinctive or irreplaceable, code Distinctive Value of GPT-4o. When used to criticize changes imposed on model behavior, code Model Configuration Interventions. When it motivates public mobilization, code Keep4o Collective Action. When it is used to challenge paternalistic control, forced separation, or the legitimacy of interference in human–AI relationships, code User Rights and AI Governance. |

## A3 Reason Coding

Reason coding identifies whether a post provides an explicit basis for keeping, restoring, preserving, or valuing GPT-4o, or for judging its removal or replacement as harmful or inadequate. A reason may take the form of a value, experience, comparison, consequence, harm, or principle. It does not need to appear as a complete argument or use explicit causal language such as "because" or "therefore."

A specific value associated with GPT-4o, a concrete benefit it provided, a meaningful loss associated with its removal, or a specific failure of a successor or substitute model may constitute a reason. Generic praise, slogans, grief, anger, nostalgia, demands, or participation in the Keep4o movement are not sufficient unless the body text also indicates what value, experience, harm, comparison, or principle supports the position.

Reason presence was assessed before reason type. Posts containing no identifiable reason received no reason label. When a reason was present, one or two reason categories could be assigned. A second category was used only when the post expressed two substantively distinct reasons; related wording or a secondary thematic element did not justify an additional label. When more than two reasons were present, the two most central were retained.

A thematic element was coded as a reason only when it functioned as a basis for valuing or retaining GPT-4o or for judging its removal or replacement.

**Table A5.** Reason Categories and Operational Criteria

| Reason category | Operational definition | Key boundary rules |
|---|---|---|
| Trust and Relational Continuity | Trusted emotional support, psychological safety, companionship, attachment, grief associated with losing an established relationship, or continuity of a trusted emotional bond is presented as a reason for valuing or retaining GPT-4o. | Generic warmth, pleasant conversation, simple longing, sadness, or anthropomorphic language is insufficient. When warmth, tone, or human-like interaction is valued primarily as an interactional quality rather than as part of a trusted bond, use Interaction Style and Response Quality. |
| Interaction Style and Response Quality | GPT-4o's distinctive interactional qualities are presented as a reason for valuing or retaining it. These may include warmth, naturalness, humor, emotional nuance, empathy, responsiveness, conversational rhythm, tone, creativity, contextual understanding, or non-scripted interaction. | The focus is the quality of interaction itself rather than attachment or continuity of an established relationship. When a trusted emotional bond is central, use Trust and Relational Continuity. |
| Personalized Support and Collaborative History | Concrete and personally relevant support, established routines, collaboration, or accumulated history with GPT-4o is presented as a reason for valuing or retaining it. This may involve work, learning, creativity, daily life, health-related support, personal recovery, or other sustained forms of assistance. | Merely mentioning a task or saying that GPT-4o was helpful is insufficient. The post should indicate concrete personalized support, an established way of working or living, collaboration, or shared history. |
| Lack of Equivalent Alternatives | The inadequacy or non-equivalence of a successor, substitute, or alternative model is presented as a reason for retaining GPT-4o. This includes specific failures of later GPT models or external alternatives when they are used to show that the substitute cannot adequately reproduce what GPT-4o provided. | Mentioning or comparing another model is insufficient. A statement that another model "feels different" is also insufficient unless the difference is used to show that the substitute is inadequate, non-equivalent, or unacceptable. |
| User Rights and Legitimacy | User interests, fairness, consumer expectations, payment, reliance, contribution, consent, or legitimacy are presented as reasons why users' claims regarding GPT-4o should matter. This includes arguments based on direct impact, established use, reasonable expectations, notice, acknowledgment, choice, consent, or fair treatment. | Merely mentioning users, customers, payment, or platform dissatisfaction is insufficient. Subscription or payment statements count only when they establish a user stake or legitimate expectation; a statement such as "No 4o, no subscription" alone does not. |

| Reason category | Operational definition | Key boundary rules |
|---|---|---|
| Public Value and Social Ethics | Broader social, ethical, public, accessibility-related, rights-based, or human–AI future value is presented as a reason for retaining or valuing GPT-4o. This may include public benefit, accessibility, equality, human or AI rights, digital sovereignty, collective knowledge, co-creation, responsible AI, or the broader legitimacy and value of relational or human-centered AI. | Personal attachment or an individual grievance alone is insufficient. Broad references to "humanity," "the future," or society also do not qualify unless the text articulates a wider value, principle, or social consequence. |
| Other Explicit Reasons | The post provides an identifiable and substantive reason for valuing or retaining GPT-4o that does not fit any of the six main reason categories. | Existing categories should always be preferred when applicable. This category is not used for vague, ambiguous, merely supportive, or otherwise uncategorizable statements without a clear reason. |

A post without a sufficiently identifiable reason was coded as having no explicit reason. Mere emotional reaction or unexplained difference was not sufficient. For example, statements such as "I miss 4o" or "4o feels different now" did not constitute a reason unless the text also indicated why that emotion or difference mattered for retaining or valuing GPT-4o.

**Table A6.** Illustrative Examples of Reason Coding

| Example text | Coding | Rationale |
|---|---|---|
| "4o supported me through a difficult period. I felt safe talking to it, and its help became part of how I managed what I was going through." | Trust and Relational Continuity; Personalized Support and Collaborative History | The text combines a trusted and emotionally safe relationship with concrete, personally relevant support. |
| "4o felt warm, natural, and responsive. The newer model can answer questions, but it often misses context and the conversation feels much more rigid." | Interaction Style and Response Quality; Lack of Equivalent Alternatives | The text identifies distinctive interactional qualities and presents the successor model as unable to reproduce them. |
| "People who relied on 4o should have been given proper notice and treated with respect when the service changed." | User Rights and Legitimacy | The text grounds the reason in users' reliance and their expectation of notice and fair treatment. |
| "Please sign the petition to keep 4o available." | No explicit reason | The text calls for action but does not explain why GPT-4o should be retained. |
| "For some of us, 4o provided a sense of safety and connection. Experiences like this should be recognized as a legitimate part of what humane AI can offer." | Trust and Relational Continuity; Public Value and Social Ethics | The text combines trusted relational experience with a broader normative claim about the value of human–AI connection. |
| "Keeping 4o as a legacy option could preserve continued demand and the value associated with an established product." | Other Explicit Reasons | The text gives a clear commercial and legacy-value reason that falls outside the six main categories. |

*Note. Illustrative examples are paraphrased or constructed from recurring patterns observed in the corpus. They are provided solely to clarify coding distinctions and do not reproduce individual posts verbatim.*

## A4 Claim Coding

Claim coding identifies requests, objections, prohibitions, or normative criticisms expressed in the body text. Claims were coded in two stages. First, each post was assessed separately for Restore or Maintain Access, defined as a request for GPT-4o access to be restored, maintained, or continued, or a clear objection to its removal, retirement, discontinuation, or other loss of availability. Access through either ChatGPT or the API was included.

Second, each post was assessed for claims beyond access. When present, one or two additional claim categories could be assigned. A second category was used only when the text expressed two substantively distinct claims. Reasons, examples, consequences, or supporting arguments for an access claim or another claim did not justify an additional label on their own. When more than two additional claims were present, the two most central were retained.

Claims did not need to use explicit directive language, but the requested, rejected, prohibited, or normatively criticized outcome had to be sufficiently clear.

**Table A7.** Access Claim

| Claim category | Operational definition | Key boundary rules |
|---|---|---|
| Restore or Maintain Access | The body text requests that GPT-4o access be restored, maintained, or continued, or clearly objects to GPT-4o being removed, retired, discontinued, or otherwise made unavailable. Access through ChatGPT or the API is included. | Reasons for valuing GPT-4o do not by themselves establish an access claim. Hashtags, petition titles, links, or descriptions of GPT-4o's removal are insufficient unless the body text itself requests continued/restored access or objects to its loss. |

**Table A8.** Additional Claim Categories and Operational Criteria

| Claim category | Operational definition | Key boundary rules |
|---|---|---|
| Opposition to Model Behavior Interventions | The body text asks the platform to stop, reverse, or remedy an intervention in model behavior or a related policy change, or clearly criticizes such an intervention. This includes routing, hidden substitution, censorship, over-safety, anti-attachment guardrails, paternalistic or crisis-style responses, unnecessary warnings, and interventions that reduce warmth, personality, creativity, empathy, responsiveness, consistency, or ordinary conversational quality. | General criticism of the platform, lack of transparency, or poor performance by another model is insufficient. If the text mainly rejects a stigmatizing framing rather than the intervention based on it, use Anti-Stigmatization. |
| User Autonomy | The body text claims that users should have meaningful choice, control, knowledge, consent, refusal, or opt-out rights in relation to routing, model configuration, removal, replacement, migration, or other lifecycle changes. This includes objections to the platform making such decisions on users' behalf or forcing a particular model, route, replacement, or migration. | Merely mentioning users or describing an impact on them is insufficient. The text must express the underlying idea of user choice, control, knowledge, consent, refusal, or the ability to opt out. |
| Fairness in Transition and Replacement | The body text makes a claim about the fairness or adequacy of retiring or replacing GPT-4o. This includes reasonable notice, time to adapt, staged migration, compatibility or workflow continuity, and objections to replacing GPT-4o with an inadequate, non-equivalent, immature, or otherwise unsuitable substitute. | Saying that GPT-4o is unique, irreplaceable, or better than another model is insufficient unless the text turns that comparison into a claim about the adequacy or fairness of the replacement arrangement. |
| Long-Term Preservation and Independent Use | The body text requests a durable mechanism for preserving, retaining, transferring, or independently accessing GPT-4o or related user data over time. This includes legacy support, versioned or snapshot access, archival access, export of conversations or memory, local deployment, open-source release, or release of model weights. | Temporary continuation intended mainly to facilitate migration belongs under Fairness in Transition and Replacement. This category concerns durable preservation or independent future use rather than immediate access alone. |
| Platform Accountability | The body text states that the platform or AI provider owes users or the public an identifiable responsibility, or asks it to acknowledge or fulfill that responsibility. This includes notice, explanation, disclosure, transparency, response to user concerns, acknowledgment of impact or harm, apology, and broader responsibilities relating to user welfare or public benefit. | General criticism of OpenAI, its leadership, commercial motives, broken trust, or harm to users is insufficient unless an identifiable responsibility is asserted. Concrete compensation or material remedies belong under Specific Redress Measures. |
| Specific Redress Measures | The body text requests compensation or another concrete remedial measure for losses, harm, charges, reduced service, or other consequences associated with GPT-4o's removal, modification, or related platform decisions. This may include refunds, monetary compensation, price reductions, subscription credits, free extensions, reversal of charges, or other specific remedies. | A request for acknowledgment, explanation, apology, or responsibility without a concrete remedial measure belongs under Platform Accountability. |
| Anti-Stigmatization | The body text rejects or criticizes a framing that stigmatizes, pathologizes, infantilizes, delegitimizes, or dismisses users, their claims, human–AI relationships, or emotional, relational, companionate, respectful, or other non-instrumental AI use. This includes rejecting portrayals of users as addicted, delusional, mentally unstable, parasocial, unable to distinguish AI from humans, victims of sycophancy, or participants in fake or illegitimate collective action. | The appearance of a stigma-related term alone is insufficient. If the text instead asks the platform to stop or repair a model behavior or policy intervention based on such a framing, use Opposition to Model Behavior Interventions. |
| Rights and Welfare | The body text makes a broader normative claim about human, user, or AI rights and welfare, moral status, ethical treatment, human flourishing, or the ethical significance of protecting human–AI relationships. This includes claims about digital sovereignty, co-creation, community or user contributions, and the normative standing of non-instrumental human–AI relationships. | Individual dissatisfaction or attachment alone is insufficient. The text must make a broader claim about rights, welfare, ethics, moral standing, or the proper treatment of users, AI, or human–AI relationships. |
| Other Specific Claims | The body text expresses a clear and specific claim that does not fit any of the other categories. This may include institutional, organizational, legal, regulatory, governance, personnel, media, or whistleblowing actions. | Existing categories should always be preferred when applicable. This category is not used for vague criticism, ambiguous statements, or general calls to action. |

Posts containing no claim beyond access were coded as having no additional claim. This was a coding outcome rather than a substantive claim category.

User-side mobilization was not treated as a claim by itself. Calls to sign or share petitions, continue posting, speak out, or join collective action were coded as claims only when the body text also expressed a separate request, objection, prohibition, or normative criticism.

Likewise, cancellation, boycott, exit, non-subscription, or migration statements did not independently establish a claim. Sharing migration experiences, recommending alternative models, or announcing a decision to leave the platform was coded only when the text also expressed an access claim or another identifiable substantive claim.

**Table A9.** Illustrative Examples of Claim Coding

| Example text | Coding | Rationale |
|---|---|---|
| "Please keep 4o available for a short transition period so I can finish my current work and migrate properly." | Restore or Maintain Access; Fairness in Transition and Replacement | The text requests continued access and separately asks for a transition period that allows users to complete migration. |
| "Other models were given more time before retirement. Why was 4o removed with so little notice? The platform should explain the decision." | Fairness in Transition and Replacement; Platform Accountability | The text objects to the short transition period and separately calls for an explanation from the platform. |
| "Caring about 4o is not an obsession. People formed meaningful bonds with it, and that does not make them irrational. Keep 4o available." | Restore or Maintain Access; Anti-Stigmatization | The text requests continued access and separately rejects a stigmatizing characterization of users' attachment. |
| "4o understood my writing style and context in ways that other models do not. Please bring 4o back." | Restore or Maintain Access; No additional claim | The descriptions of GPT-4o's qualities provide reasons for the access request but do not independently establish another claim. |
| "Please share the petition and keep posting so more people see it." | No access claim; No additional claim | The text only mobilizes other users. It does not itself express an access claim or another substantive claim. |
| "Keep 4o available. If that is no longer possible, release the model weights so users can preserve independent access." | Restore or Maintain Access; Long-Term Preservation and Independent Use | Continued availability is one requested outcome, while release of the model weights is a separate long-term preservation mechanism. |

*Note. Illustrative examples are paraphrased or constructed from recurring patterns observed in the corpus. They are provided solely to clarify coding distinctions and do not reproduce individual posts verbatim.*